\documentclass[aps,prl,shownopacs,superscriptaddress,twocolumn]{revtex4}

\usepackage{graphicx}
\usepackage{hyperref}
\usepackage{amsmath}
\usepackage{amssymb}
\usepackage{stmaryrd}
\usepackage{epstopdf}
\usepackage{bbm}
\usepackage{pgfplotstable}
\usepackage{physics}
\usepackage{bbold}
\usepackage{gensymb}
\usepackage{siunitx}

\begin{document}

\title{Remote preparation of continuous-variable qubits\\ using loss-tolerant hybrid entanglement of light}

\author{H. Le Jeannic}
\affiliation{Laboratoire Kastler Brossel, Sorbonne Universit\'e, CNRS, ENS-Universit\'e PSL, Coll\`ege de France, 4 Place
Jussieu, 75005 Paris, France}
\author{A. Cavaill\`{e}s}
\affiliation{Laboratoire Kastler Brossel, Sorbonne Universit\'e, CNRS, ENS-Universit\'e PSL, Coll\`ege de France, 4 Place
Jussieu, 75005 Paris, France}
\author{J. Raskop}
\affiliation{Laboratoire Kastler Brossel, Sorbonne Universit\'e, CNRS, ENS-Universit\'e PSL, Coll\`ege de France, 4 Place
Jussieu, 75005 Paris, France}
\author{K. Huang}
\affiliation{Shanghai Key Laboratory of Modern Optical Systems and Engineering Research Center of Optical Instruments and Systems (Ministry of Education), School of Optical Electrical and Computer Engineering, University of Shanghai for Science and Technology, Shanghai 200093, China}
\author{J. Laurat}
\email{julien.laurat@sorbonne-universite.fr}
\affiliation{Laboratoire Kastler Brossel, Sorbonne Universit\'e, CNRS, ENS-Universit\'e PSL, Coll\`ege de France, 4 Place
Jussieu, 75005 Paris, France}

\begin{abstract}
Transferring quantum information between distant nodes of a network is a key capability. This transfer can be realized via remote state preparation where two parties share entanglement and the sender has full knowledge of the state to be communicated. Here we demonstrate such a process between heterogeneous nodes functioning with different information encodings, i.e., particle-like discrete-variable optical qubits and wave-like continuous-variable ones. Using hybrid entanglement of light as a shared resource, we prepare arbitrary coherent-state superpositions controlled by measurements on the distant discrete-encoded node. The remotely prepared states are fully characterized by quantum state tomography and negative Wigner functions are obtained. This work demonstrates a novel capability to bridge discrete- and continuous-variable platforms.
\end{abstract}

\maketitle

In the context of quantum networks, remote state preparation (RSP) protocols enable the transfer of quantum information from one place to a distant one via entanglement shared between two parties \cite{Lo2000,Pati2000,Bennett2001,Berry2003}. In contrast with quantum teleportation \cite{reviewteleport}, the sender has complete knowledge of the state to be communicated. Conditioned on the sender's measurement and one-way classical communication, the receiver's state is projected onto the targeted state. RSP finds a variety of applications, ranging from long-distance quantum communication to loss-tolerant quantum-enhanced metrology \cite{FurusawaRev}.

In recent years, a number of demonstrations have been realized. Remote state preparations of polarization qubits were demonstrated based on polarization entanglement \cite{Peters,Xiang2005,Liu2007}. Transfer of single-photon and vacuum superpositions was also achieved based on single-photon entanglement \cite{Babichev2004} and continuous-variable RSP was demonstrated using Einstein-Podolsky-Rosen entangled beams \cite{Paris2003a,Laurat2003,Kurucz2005}. These works were extended to the preparation of multiqubit states \cite{multi2010,multi2016}, spatial qubits \cite{spatial} and single-plasmon states \cite{plasmon}. Remote preparation of atomic memories also enabled the transfer of a given state to a long-lived matter system \cite{deRiedmatten2006,Rosenfeld2007}. In all these realizations, the initial entangled resource is either based on finite-dimensional systems, such as single-photon qubits, or on infinite-dimensional spaces, such as squeezed states, following thereby the traditional separation between quantum information approaches.  

However, a large effort has been recently devoted to bridge these different approaches, i.e., discrete-variable (DV) and continuous-variable (CV) encodings and toolboxes \cite{Andersen2015}. Deterministic CV teleportation of discrete qubits has been demonstrated \cite{Furusawa} and novel hybrid protocols, such as a single-photon entanglement witness based on quadrature measurements \cite{Morin}, have been implemented. In this context, the recent demonstrations of hybrid entanglement between CV and DV optical qubits \cite{Morin2014, Jeong2014,Lvovsky2017} holds the promise of heterogenous networks where the discrete- and continuous-variable operations could be efficiently combined.

In this paper, we demonstrate a first remote preparation scheme between two distant network nodes that rely on different information encodings, i.e. discrete and continuous variables. Starting from heralded hybrid entanglement of light shared between the two nodes, arbitrary superpositions of coherent states, i.e., CV qubits, are prepared by a measurement performed on the DV node. The prepared states are then characterized by full quantum state tomography and compared to the targeted states. We detail the phase space evolution of these transferred states as a function of the triggering measurements.  

\begin{figure}[t!]
\centering
\includegraphics[width=0.93\columnwidth]{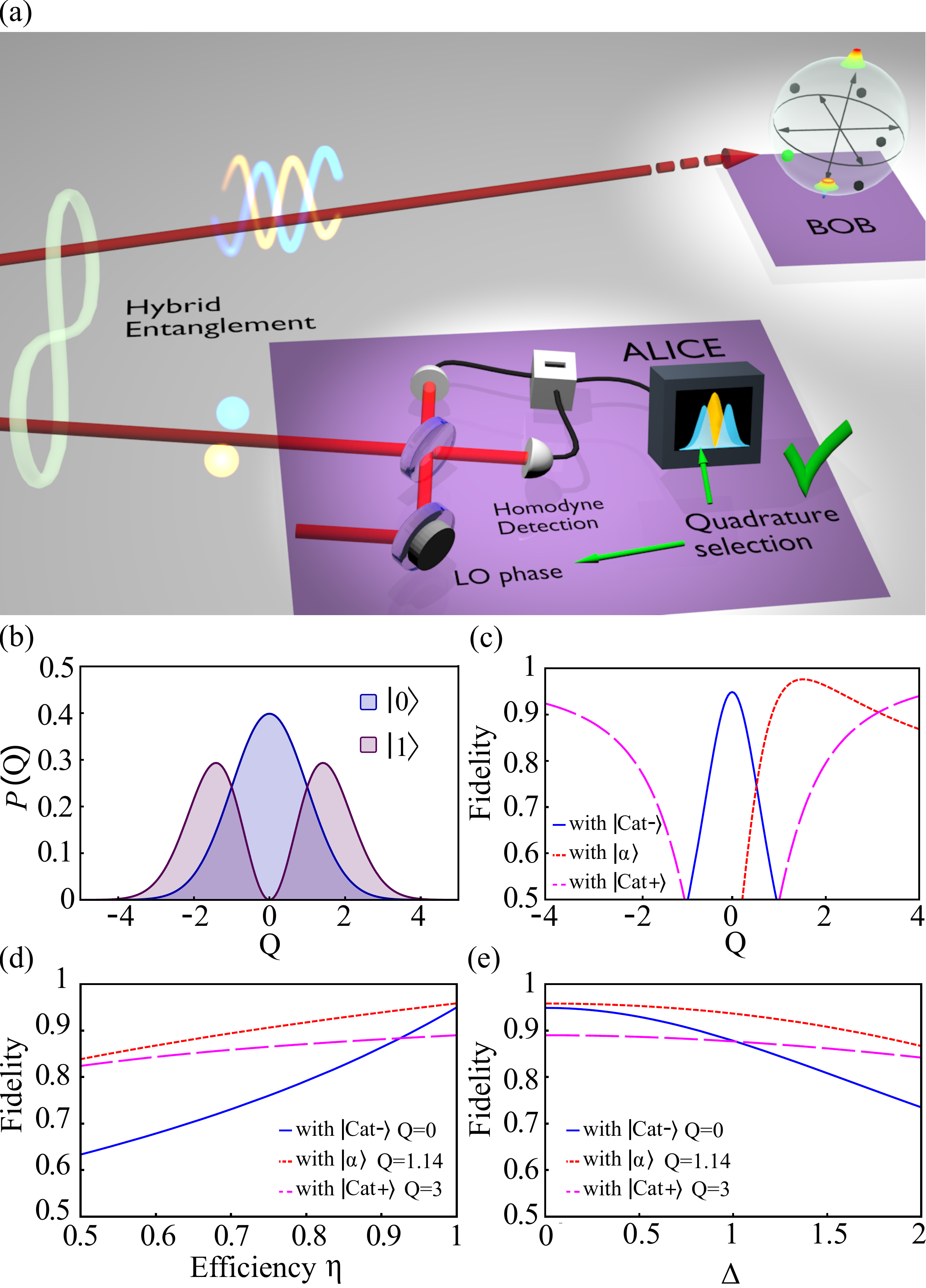}
\caption{Remote preparation of coherent-state superpositions using hybrid entanglement of light. (a) Alice and Bob located at two distant nodes share an entangled state $|0\rangle_A | \textrm{Cat}- \rangle_B + |1\rangle_A |\textrm{Cat}+\rangle_B$. Conditional on a specific quadrature measurement via homodyne detection, Alice remotely prepares any arbitrary superposition $c_+| \textrm{Cat} + \rangle +e^{i\varphi} c_- |\textrm{Cat}-\rangle$ on Bob's node. The measured quadrature is chosen by locking the local oscillator phase on a value $\theta$ and the preparation is heralded by the occurrence of a preselected value $Q$ within an acceptance window of width $\Delta$ (typically taken equal to 20\% of the shot noise value). The prepared state is characterized by homodyne detection, with an overall efficiency $\eta=85$\%. (b) Marginal distributions of vacuum and single-photon states. (c) Theoretical fidelity of the remotely prepared state with different targeted superpositions as a function of the quadrature value $Q$, with $\theta = 0$ and $|\alpha|=0.7$. (d) Theoretical fidelity as a function of the efficiency of the heralding homodyne detection. (e) Theoretical fidelity as a function of the window width $\Delta$.}
\label{figure1}
\end{figure}

The experimental setup is sketched on Fig. \ref{figure1}. It relies on the hybrid entanglement between particle-like and wave-like optical qubits we recently demonstrated \cite{Morin2014}. The entanglement is generated between two remote nodes that relied each on an optical parametric oscillator (OPO), and connected by two lossy channels. The detection of a single photon in a middle station via a superconducting nanowire single-photon detector \cite{LeJeannic2016} heralds the generation. Importantly, this measurement-induced scheme preserves the fidelity of the entangled state independently of the loss between the two nodes. This loss-tolerant feature is central to our realization and constitutes a prerequisite to obtain high-fidelity quantum states between remote locations. In our implementation, the entanglement heralding rate is 200 kHz and the overall loss in the conditioning path reaches 30 \%. Experimental details have been provided elsewhere \cite{Morin2014}.

Specifically, Alice and Bob share the hybrid entangled state:
\begin{equation}
|\Psi\rangle_{AB}\propto |0\rangle_A | \textrm{Cat}- \rangle_B + |1\rangle_A |\textrm{Cat}+\rangle_B.
\end{equation}
The state $|\textrm{Cat}- \rangle\propto|\alpha \rangle - |-\alpha \rangle \sim \hat{a}\hat{S}|0\rangle$ denotes an odd coherent-state superposition (CSS), which is approximated experimentally here by a single-photon-subtracted squeezed vacuum, while $|\textrm{Cat}+\rangle\propto|\alpha \rangle + |-\alpha \rangle\sim\hat{S}|0\rangle$ refers to an even CSS approximated by a 3-dB squeezed vacuum. Given this shared resource, Alice can measure her DV state and project Bob's state into any arbitrary coherent-state superposition.     

As Alice's mode is contained in the qubit subspace spanned only by zero and one photon Fock states (to limit the multiphoton component below 2\% for the DV side, the OPO is pumped about 100 times below threshold), she can use a quadrature measurement via an efficient homodyne detection to discriminate between the possible DV states \cite{Babichev2004}. This method enables to project onto any superposition of zero and one photon. To understand the general idea of this method, the marginal distributions of the quadrature corresponding to these states are plotted in Fig. \ref{figure1}(b) in the ideal case. The measurement of a quadrature outcome equal to zero on Alice's side has to come from the vacuum component and will therefore project Bob's state onto the state $| \textrm{Cat} - \rangle$. Similarly, a large value quadrature result, which most likely comes from the single-photon component, will project Bob's state onto the state $| \textrm{Cat} + \rangle$. By choosing a given phase $\theta$ and a quadrature value $Q$, Alice can then remotely prepare any superposition of the form $c_+| \textrm{Cat} + \rangle +e^{i\varphi} c_- |\textrm{Cat}-\rangle$. In the ideal case and large $|\alpha|$ values, a quadrature measurement equal to $+1$ projects the state onto the equally-weighted superposition $| \textrm{Cat} + \rangle+| \textrm{Cat} - \rangle\sim |\alpha\rangle$, i.e., a coherent state.

The superposition coefficients can be calculated as follows. The measurement implemented by Alice can be written in the form of the quadrature operator $\hat{Q}_{\theta}=\hat{X} \cos \theta+ \hat{P} \sin \theta$
where $\hat{X}$ and $\hat{P}$ denote the canonical position and momentum observables. The measurement of a quadrature value $Q$ projects the entangled state onto a quadrature eigenstate $\langle Q_{\theta}|$:
\begin{eqnarray}
|\Phi\rangle_{B}\propto\langle Q_{\theta}|\Psi\rangle_{AB}= \langle Q_{\theta}|0\rangle_A | \textrm{Cat} - \rangle_B + \langle Q_{\theta}|1\rangle_A |\textrm{Cat}+\rangle_B
\end{eqnarray}
with 
\begin{eqnarray}
 \langle Q_{\theta}|0\rangle_A=\frac{1}{(2\pi)^{1/4}}e^{-Q^2/4} \quad \textrm{and} \quad
\langle Q_{\theta}|1\rangle_A=\frac{Qe^{i\theta}}{(2\pi)^{1/4}}e^{-Q^2/4}. \nonumber
\end{eqnarray}
The remotely prepared state on Bob's side can finally be written after normalization as: 
\begin{equation}
|\Phi\rangle_{B}=\frac{1}{\sqrt{1+Q^2}}\left(| \textrm{Cat} - \rangle+Q e^{i\theta}| \textrm{Cat} + \rangle\right).
\end{equation}
It can be seen that the chosen quadrature value $Q$ changes the superposition weight, while the phase $\theta$ is directly mapped onto the relative phase of the superposition, as will be verified later. 

Figure \ref{figure1}(c) provides the expected fidelities to different targeted states as a function of the measured quadrature value $Q$. For this calculation, we consider our experimental case for which the mean photon number is limited due to the initial approximation in the entangled state generation. The fidelities are calculated for $|\alpha|=0.7$. As a result, for instance at $Q=0$, the fidelity of the prepared state to $|\textrm{Cat} - \rangle$ is equal to 95\%. For other values of Q, all superposition can be generated, and the measurement angle $\theta$ comes into play. In particular, for $Q=\pm 1.14$ and $\theta=0$, one can obtain the coherent state $|\pm \alpha \rangle$. This conditioning value $|Q|$ is slightly larger than 1 due to the limited size $|\alpha|^2$. 

Experimentally, two parameters can lead to a reduction a fidelity \cite{Paris2003}. A first one is the finite efficiency of the detection used for heralding, as shown in Fig. \ref{figure1}(d). The second one is the acceptance window in quadrature values. Indeed, measuring exact values of $Q$ would lead to a zero success probability. We therefore accept events in a certain window $[Q-\Delta/2, Q+\Delta/2]$. The selection width $\Delta$ results in a trade-off between preparation rate and fidelity. However, the reduction in fidelity is only of second order with the width (Fig. \ref{figure1}(e)). This enables us to take $\Delta$ equal to 20\% of the shot noise value. For such a selection band, centered for instance in Q=0, the  success rate is around 5\% while the fidelity is only decreased by a few percents. 

\begin{figure}[t!]
\centering
\includegraphics[width=0.71\columnwidth]{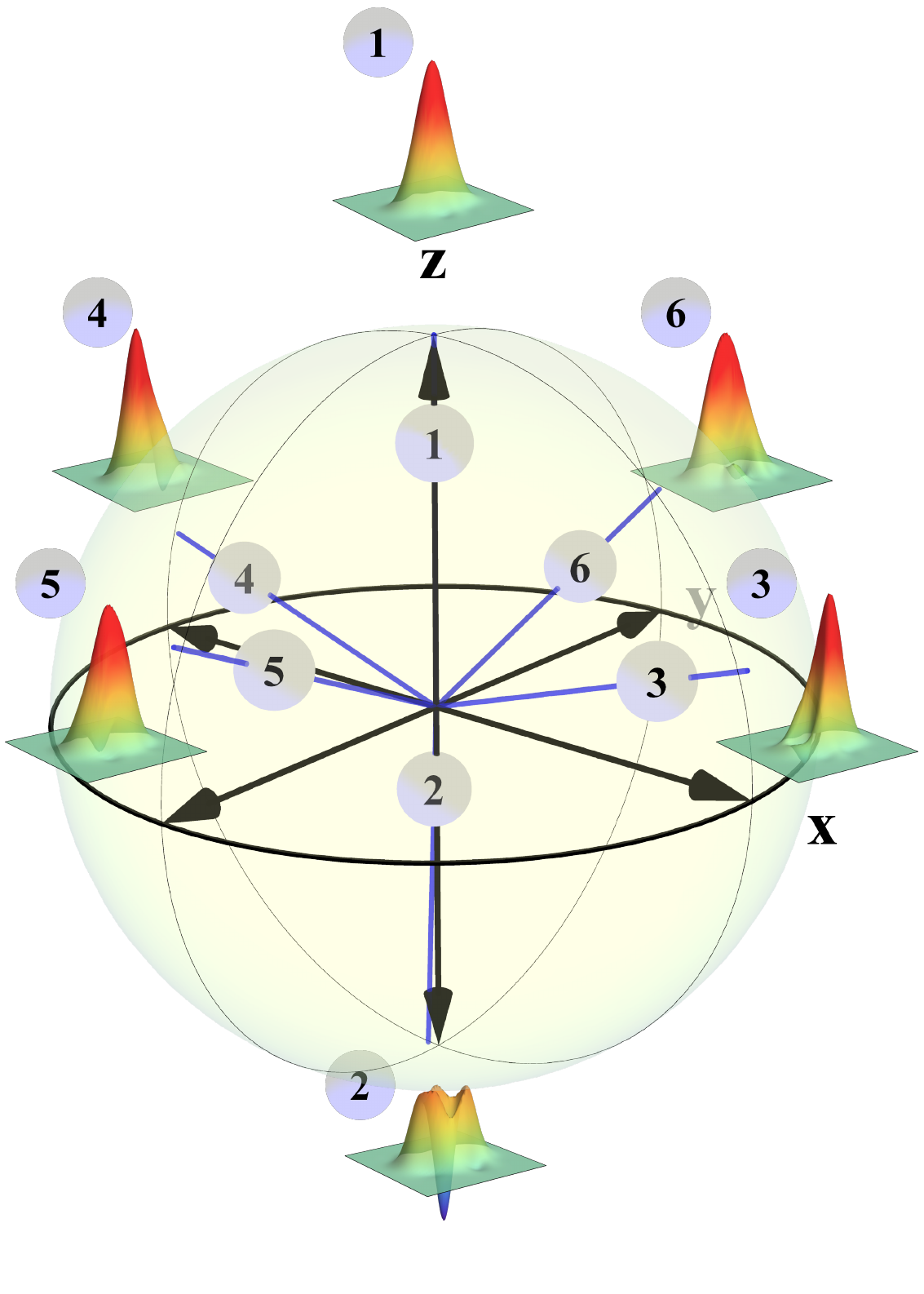}
\vspace{-0.5cm}\caption{Remotely prepared states represented on the Bloch sphere and associated Wigner functions. The poles are the two orthogonal states $|\textrm{Cat}+\rangle$ and $|\textrm{Cat}-\rangle$, with $|\alpha|=0.7$. The results are corrected for the $\eta=85$\% detection efficiency.}
\label{figure2}
\end{figure}

For a given targeted preparation, the local oscillator phase $\theta$ is locked by periodically sending a weak beam through Alice's path. The interference between this beam and the local oscillator is measured and the intensity is locked at a constant and ajustable level using a 12-bit microcontroller \cite{RSI}. The phase fluctuations are measured to be around $3^\circ$ rms. The state remotely prepared at Bob's node is characterized by quantum state tomography via homodyne detection, with a detection efficiency $\eta=85$\%. Quadrature values from 30 000 to 50 000 realizations depending on the  conditioning are recorded, and then processed via a maximum likelihood algorithm to obtain the density matrix and the associated Wigner function \cite{Tomo,MorinTomo}. 

We come to the experimental results. A set of remotely-prepared states are presented in Fig. \ref{figure2}, inserted in a Bloch sphere where the poles are defined by the orthogonal states $|\textrm{Cat}+\rangle$ and $|\textrm{Cat}-\rangle$, with $|\alpha|=0.7$. In order to graphically represent the states, for each prepared state $\hat{\rho}_{\textrm{exp}}$ we determine the maximal fidelity with the state $\cos (\phi/2)|\textrm{Cat}~+~\rangle+ ~e^{i\varphi}~ \sin (\phi/2) |\textrm{Cat}-\rangle$
and obtain thereby the spherical coordinates $\{\phi,\varphi \}$. The distance $d$ to the center scales with the purity of the state, $d~=~\sqrt{2\textrm{Tr}[\hat{\rho}_{\textrm{exp}}^2]-1}$. Each prepared state is represented by a number located in the sphere and we give next to it the corresponding experimental Wigner function. Table \ref{Table_RSP} gives a summary of the conditioning parameters used to prepare these states and the fidelity with the targeted state to be transferred.

\begin{table}[t!]
\small
\centering
\begin{tabular}{|lllll|l}
\hline
$\#$ & Target \quad&Q, $\theta$ & $\mathcal{F}_{|\alpha|=0.7}$ & Rate\\
\hline 
1 & $ |\textrm{Cat}+\rangle$& $|Q|\geq 2$, $\theta=0$&$86\%$& 13.8 kHz\\
2 & $ |\textrm{Cat}-\rangle$& $Q=0$, $\theta=0$&$65\%$& 9.6 kHz\\
3 & $| \alpha \rangle$& $Q=1.14$, $\theta=0$& $85\%$& 9.4 kHz \\
4 & $|-\alpha \rangle$& $Q=-1.14$, $\theta=0$& $85\%$& 9.4 kHz \\
5 & $| \alpha \rangle+i|-\alpha \rangle$& $Q=-1.14$, $\theta=\pi/2$& $81\%$& 9.4 kHz \\
6 & $| \alpha \rangle-i|-\alpha \rangle$& $Q=1.14$, $\theta=\pi/2$&$80\%$ & 9.4 kHz\\
\hline
\end{tabular}
\caption{Summary of the prepared states corresponding to each point on Fig. \ref{figure2}. The targeted states appear in the second column as well as the experimental fidelities $\mathcal{F}$ for $|\alpha|=0.7$. $Q$ and $\theta$ correspond to the quadrature value in unit of shot noise and to the local oscillator phase. For the points 2 to 6, the acceptance window $\Delta$ is equal to 0.2. The error bar on the fidelity is $\pm 3\%$. The last column provides the heralded rate.}
\label{Table_RSP}
\end{table}

\begin{figure*}[t]
\centering
\includegraphics[width=1.8\columnwidth]{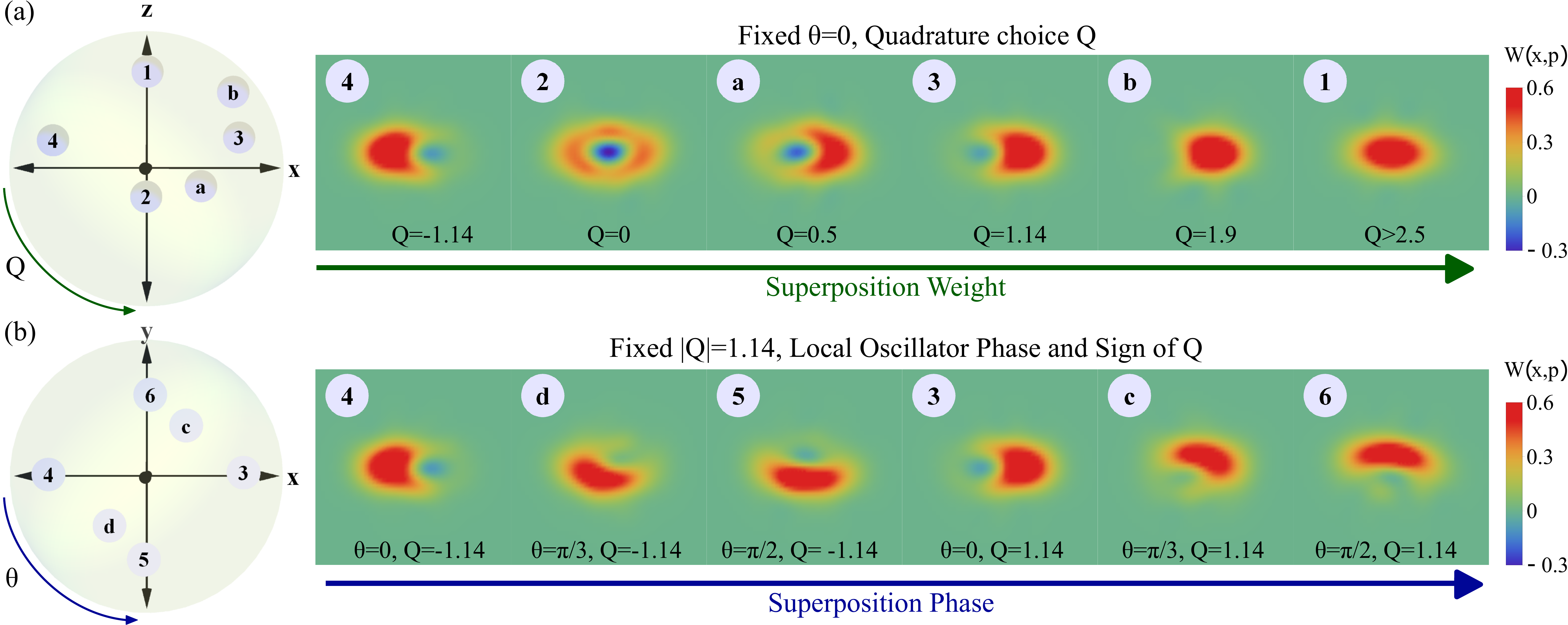}
\caption{Control of the remotely-prepared superposition via tuning of (a) the selected quadrature value $Q$ and (b) the local oscillator phase $\theta$. The left figures provide the location of the prepared states projected onto the XZ and XY plane of the Bloch sphere respectively. The rows show the evolutions of the associated Wigner functions when the conditioning parameter is tuned. We note that the sign flip of $Q$ is equivalent to a $\pi$ phase shift. The results are corrected for the $\eta=85$\% detection efficiency.}
\label{figure3}
\end{figure*}

As these results show, our procedure enables the remote preparation of arbitrary CV qubits with a large fidelity to the target states. Fidelities above 80\% are obtained, except for the state numbered 2, i.e., for a $|\textrm{Cat}-\rangle$ target. This reduced fidelity is generally true for states lying closer to the south pole. Indeed they are obtained using a measurement close to $Q=0$. The conditioning has therefore a non-negligible probability to come from the initial component $|1 \rangle_A \hat{S}|0\rangle_B $ that has experienced photon losses on Alice's side, where no loss correction can be applied. This is in contrast to large $Q$ values. We note also that the states numbered 3 and 4 are close to coherent states with opposite phase, as expected. However, they are not exactly lying on the sphere equator but slightly out of this plane due to the limited mean photon number $|\alpha|^2<1$. 

We now investigate more in detail the control of the prepared superpositions as a function of the conditioning parameters. Figure \ref{figure3} provides projections of the Bloch sphere along two planes, i.e., XZ and XY. Rotation of the states in these planes are controlled by independent parameters, namely the quadrature value $Q$ and the phase $\theta$, respectively. While the first one changes the weight of the superposition, the second one modifies its relative phase. This result illustrates the quantum state engineering capability in phase space offered by the present scheme.

The performance of our procedure is currently limited by the mean-photon number of our transferred states, with $|\alpha|^2\sim0.5$. By enhancing the detection efficiency on Alice's side, which is mainly reduced by an optical isolator used to avoid any backscattering from the detection system, it is possible to directly increase this size to $|\alpha|^2\sim1$. Extensions of the entanglement scheme to the recently demonstrated techniques for large optical cat state generation \cite{Etesse2015,Huang2015} would enable to reach values above 2, for which the overlap between the coherent-state components drops below $10^{-3}$ and enables fault-tolerant operations \cite{Lund2008}.

In summary, we have described a RSP experiment based on hybrid entanglement of light generated by a measurement-induced protocol. This scheme first enables the challenging quantum state engineering at a distance of non-Gaussian states that are vulnerable to losses. It also makes possible the interfacing of distant quantum nodes based on different encodings and therefore the exchange of quantum information in a heterogeneous network. Within the broad setting of the optical hybrid approach to quantum information, this work paves the way towards the demonstration of EPR steering and the investigation of semi-device-independent communication scenarios.

\begin{acknowledgments}
The authors thank O. Morin and Y. Hashimoto for their contributions in the early stage of the experiment. This work was supported by the European Research Council (Starting Grant HybridNet), Sorbonne Universit\'e (PERSU program) and the French National Research Agency (Hy-Light project). K.H. acknowledges the Program for Professor of Special Appointment (Eastern Scholar) at Shanghai Institutions of Higher Learning, and Science and Technology Innovation Program of Basic Science Foundation of Shanghai (18JC1412000).
\end{acknowledgments}

\end{document}